\documentclass[prd, twocolumn,nofootinbib,notitlepage,floatfix]{revtex4-1}
\pdfoutput=1

\usepackage{amsmath}
\usepackage{graphicx}
\usepackage{dcolumn}
\usepackage{bm}
\usepackage{epsfig}
\usepackage{amssymb,latexsym,mathrsfs}
\usepackage{graphicx}
\usepackage{color}
\usepackage{hyperref}

\hypersetup{
    colorlinks=true,
    linkcolor=red,
    citecolor=blue,
} 

\newcommand{\be}{\begin{equation}}
\newcommand{\ee}{\end{equation}}
\newcommand{\bs}{\begin{split}} 
\newcommand{\bea}{\begin{eqnarray}}
\newcommand{\eea}{\end{eqnarray}}

\newcommand{\om}{\Omega_m}

\newcommand{\ode}{\Omega_{\rm de}}
\newcommand{\lcdm}{$\Lambda$CDM}

\newcommand{\zl}{z_{\rm lens}} 
\newcommand{\zs}{z_{\rm source}}

\begin{document}

\title{Double Source Lensing Probing High Redshift Cosmology} 

\author{Divij Sharma$^1$, Eric V.~Linder$^{2,3}$}
\affiliation{
$^1$Department of Physics, University of California, Berkeley, CA 94720, USA\\  
$^2$Berkeley Center for Cosmological Physics \& Berkeley Lab, 
University of California, Berkeley, CA 94720, USA\\  
$^3$Energetic Cosmos Laboratory, Nazarbayev University, 
Nur-Sultan, 010000, Kazakhstan } 
\email{divijsharma@berkeley.edu, evlinder@lbl.gov}

\begin{abstract} 
Double source lensing, with two sources lensed by the same 
foreground galaxy, involves the distance between each source 
and the lens and hence is a probe of the universe away from 
the observer. The double source distance ratio also reduces sensitivity to 
the lens model and has good complementarity with standard 
distance probes. We show that using this technique at high 
redshifts $z>1$, to be enabled by data from the Euclid satellite 
and other surveys, can give insights on 
dark energy, both in terms of $w_0$--$w_a$ and redshift binned 
density. 
We find a dark energy figure of merit of 245 from 
combination of 256 double source systems with moderate 
quality cosmic microwave background and supernova data. 
Using instead five redshift bins between $z=1.1$--5, 
we could detect the dark energy density out to $z\approx5$, or make measurements ranging between 31$\sigma$ and 2.5$\sigma$ 
of its values in the bins. 
\end{abstract}

\date{\today} 

\maketitle


\section{Introduction}

Gravitational lensing gives a visual manifestation of general 
relativity in the universe, deflecting light from distance 
sources by foreground mass concentrations. When the lensing is 
strong, multiple images occur, with angular separations determined 
by the Einstein radius combining the lens mass with a ``focal 
length'' involving distances between the source, lens, and observer. 
In the particular situation of multiple sources lensed by the 
same mass, generally known as double source plane lensing (DSPL), 
the ratio of Einstein radii or deflection angles measured by image separations involves 
a pure distance ratio; the impact of the lens mass profile details 
-- modeling this can be a significant source of uncertainty in 
strong lensing -- is much reduced \cite{collett12,collett14} 
(also see \cite{schneider}). 

Moreover, the key distance ratio is a purely geometric probe, 
reflecting the cosmic expansion history separate from the growth 
history uncertainties, and involves distances between the source 
and lens, removed from the observer, i.e.\ probing the distant universe separated from the local 
universe. Furthermore it is independent 
of the Hubble constant $H_0$. This offers the interesting possibility of exploring 
the Hubble parameter and matter-energy contents of the universe 
at redshifts far from the observer, 
and with different covariances than 
other distances. 

Previous studies of the cosmological leverage of image separations 
and DSPL \cite{linder04,collett12,collett14,linder16} showed useful complementarity 
with other lensing probes, strengthening their dark energy 
figure of merit by $\sim40\%$. Those investigations focused on 
lens redshifts $z\le0.6$. Here we consider higher redshift systems, 
as will be enabled soon by the Euclid satellite \cite{euclid1,euclid2,euclid3}, 
and later by proposed higher redshift surveys such as MegaMapper 
\cite{megamapper} and others \cite{snowhi}. We also 
explore complementarity with standard distance probes, 
and go beyond the standard 
dark energy equation of state redshift dependence and allow 
dark energy density to vary freely in bins of redshift, testing 
for ``early'' dark energy behavior at $z\approx1$--5. 

One could also go beyond galaxy-galaxy-galaxy (one lens, 
two sources) lensing to use the cosmic microwave 
background (CMB) as a source plane \cite{1605.05337}, 
or go beyond probing expansion history to look at 
effects of modified gravitational potentials on the 
deflection angles \cite{1906.06324}, 
although we do not address those here. 

In Section~\ref{sec:sens} we investigate the sensitivity of  
the DSPL distance ratio to cosmological parameters, showing that 
it exhibits unique properties relative to other distance probes. 
Section~\ref{sec:constraints} propagates this to projected 
parameter estimation uncertainties, for various redshift ranges 
of observations and combinations with other data. 
In Section~\ref{sec:sourcez} we study the impact 
of the source redshift distribution. 
We explore 
early dark energy constraints in Section~\ref{sec:bins}, 
allowing independent redshift bins of dark energy density, 
and summarize, discuss, and conclude in Section~\ref{sec:concl}.

\section{Cosmological Sensitivity of DSPL} \label{sec:sens} 

The critical surface mass density for strong lensing involves the distance ratio 
$r_s/(r_l r_{ls})$ of distances between source and observer, 
lens and observer, and source and lens, respectively. The 
ratio of light deflection angles (or Einstein radii when the 
lens mass factor cancels out) for two sources with a common 
lens is the ratio of distance ratios, and the 
central quantity for DSPL, 
\bea  
\beta(z,z_1,z_2)&\equiv&\frac{r_{ls}(z,z_1)}{r_s(z_1)}\frac{r_s(z_2)}{r_{ls}(z,z_2)}\\ 
&=&\frac{D_{ls}(z,z_1)}{D_s(z_1)}\frac{D_s(z_2)}{D_{ls}(z,z_2)}\ .
\eea  
where the lens is at redshift $z$, the nearer source is at $z_1$, 
and the further source at $z_2$. Here $r(z,z_i)$ is the angular 
distance to redshift $z_i$ seen by an observer at redshift $z$, 
with the single argument $r(z_i)$ indicating the distance is 
measured from redshift zero, and 
\be 
D(z,z_i)=\int_z^{z_i} \frac{dz'}{H(z')}\,  
\ee 
is the conformal distance for a flat universe as we will use. 
Note that $\beta$ is the same whether using angular or 
conformal distances, as long as all the distances are 
treated consistently. Conformal distances in a flat universe 
have the convenient property that $D(z,z_i)=D(z_i)-D(z)$. 

We consider measurements of $\beta$ from observed image positions 
of strong lensing systems. Since Einstein radii involve lens 
mass factors as well, the ratio formed from image separations 
or positions is not strictly a function of distance only, 
except in special cases like a point mass or singular isothermal 
sphere lens profile. 
However, 
the dependence on lens mass 
profile (and its uncertainties, including substructure, 
and mass sheet degeneracies) is 
expected to be suppressed for DSPL relative to other uses of lensing; see \cite{collett12,collett14,23inLin2016,schneider,25inLin2016}. 
Also, we focus on galaxy lenses, where any residual mass effects could be modeled more easily. 
Nevertheless, followup high
resolution mapping (and possibly spectroscopy) of the lens will be an
important adjunct. For a few hundred lenses this should	not be a major	
observational program. 

Deep, wide field surveys such as 
that from the Euclid satellite (and less deep from the Vera 
C.\ Rubin Observatory's Legacy Survey of Space and Time (LSST \cite{lsst}) 
and less wide but higher resolution from the Nancy Grace Roman Space Telescope \cite{roman}) 
should find hundreds of DSPL. Even focusing on the best observed 
systems should deliver a data set of $\sim160$ from Euclid, 
potentially doubling when adding in other surveys \cite{23inLin2016,26inLin2016,27inLin2016,lenspop,1803.03604,oh,2010.15173}. 

We 
emphasize that we do not employ the abundance or distribution of
lensed systems as a cosmological probe, which would involve a complicated
blend of cosmology and survey characteristics and selection functions.
Rather, we use the properties of individual systems and, 
in the manner of time delay (single source plane) lenses, we
can use a data set of the best observed systems. 

\begin{figure}[h]
\centering 
\includegraphics[width=\columnwidth]{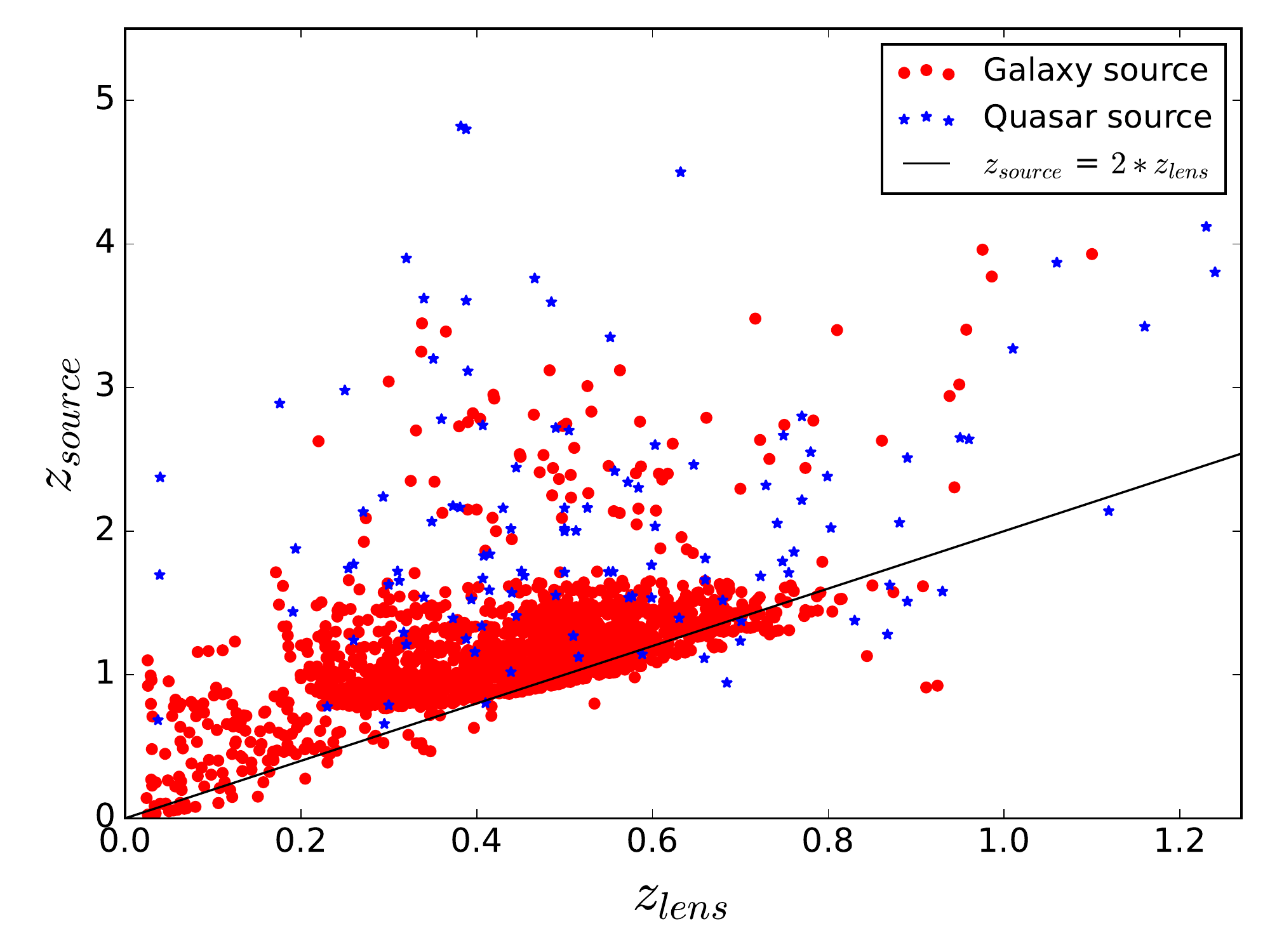}
\caption{For known galaxy lens systems, the distribution of $z_{\rm source}$ vs $\zl$ is reasonably approximated by $z_{\rm source}=2\,\zl$. Data are extracted from \cite{brownstein}. 
}
\label{fig:knownlenses} 
\end{figure}

The sensitivity of the measured $\beta$ to cosmological 
parameters is calculated through the partial derivatives, 
as a fractional change relative to some measurement precision, 
$(\partial\beta/\partial\theta)/\sigma(\beta)$. The 
cosmological parameters $\theta$ we use are $\om$, the matter 
density today as a fraction of the critical density, and initially the dark 
energy equation of state parameters $w_0$ and $w_a$, describing 
its present value and a measure of its time variation. We 
take a flat $\Lambda$CDM universe with fiducial values 
$\om=0.3$, $w_0=-1$, $w_a=0$. For illustration purposes we 
adopt a single measurement precision of 1\%, 
$\sigma(\beta)=0.01\beta$, and show $\beta(z,z_1,z_2)$ and 
its derivatives for fixed $z_1/z=2$, $z_2/z_1=1.5$. These 
ratios correspond roughly to the peak of the lensing ``focal 
length'' kernel, i.e.\ the most efficient and hence most 
commonly detected. Variations of these ratios were explored 
in \cite{linder16} and will be here as well in Section~\ref{sec:sourcez}, after the next, 
motivational paragraph. 
Moreover, 
we will find in Section~\ref{sec:sourcez}, where we vary the redshift
ratios, that this fiducial choice (basically the lower envelope
in Figure~\ref{fig:knownlenses}) will give the most conservative 
dark energy constraints
(figure of merit) -- actual data may well give more advantageous
leverage. 

While there are very few DSPL currently known, we can explore 
the reasonableness of the redshift ratio 
$z_{\rm source}/\zl\approx2$, motivated by the lensing kernel, 
for known standard strong lenses. Figure~\ref{fig:knownlenses} 
plots the redshift ratio for 1842 galaxy-galaxy and 117 quasar-galaxy strong 
lenses where the source and lens redshifts have been measured \cite{brownstein}. 
We see that indeed the value $z_{\rm source}/\zl\approx2$ is 
a reasonable approximation. In Section~\ref{sec:sourcez} we will 
quantify 
the impact on our results if we alter this.

A nice property of $\beta$ for the conditions given is that 
$\beta$ is nearly constant for a wide range of 
redshifts.  
Hence there is negligible difference between taking an 
absolute measurement precision or a fractional measurement 
precision. 
The 
1\% fiducial fractional precision for measurement of $\beta$ will depend
on survey properties, though it is likely to be a conservative
choice. For example, \cite{collett14} in 2014 achieved 1.1\% 
fractional precision
on $\beta$; the subsequent improvement in telescopes and 
instrumentation, and
the development of, e.g.\ machine learning, tools for finding and
measuring lens systems may indicate that better than 1\% will be
achieved. While we stay with the conservative choice, we note that
parameter constraints from the lensing data alone will scale linearly
with the statistical precision, while somewhat more slowly when
external data such as CMB data is included. 

Our results (from $\beta$, without other 
data) will scale with the precision. 

\begin{figure}[h]
\centering 
\includegraphics[width=\columnwidth]{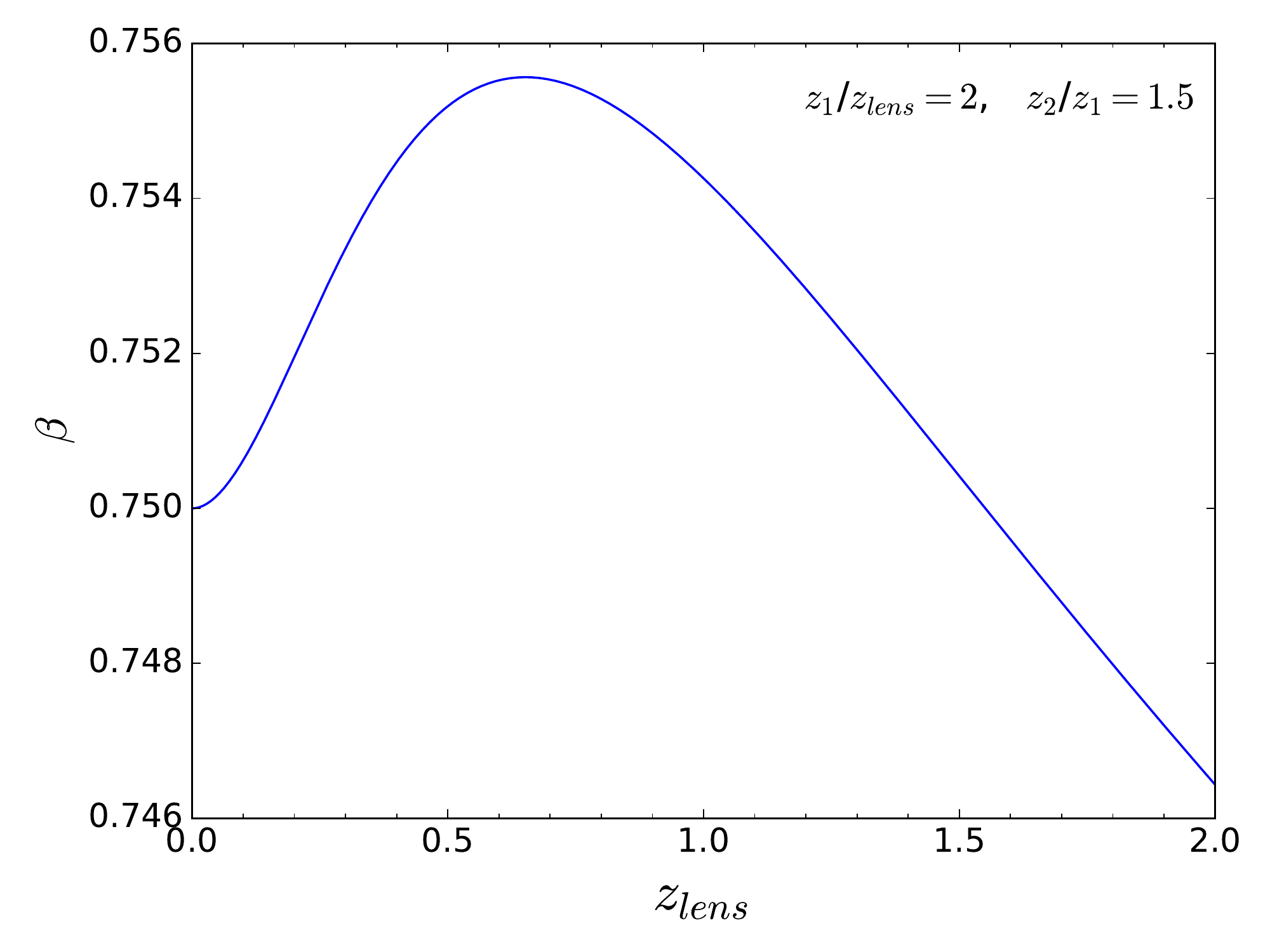}
\caption{Double source plane lensing distance ratio $\beta$ is nearly constant as a function of lens redshift $\zl$. 
} 
\label{fig:beta} 
\end{figure}

Figure~\ref{fig:beta} illustrates $\beta$ as a function 
of lens redshift, showing its near constancy, with 
deviations remaining less than 1\% out to $\zl=2$. The limit as 
$z\equiv\zl\to0$ is readily calculable as 
$\beta\to (1-z/z_1)/(1-z/z_2)$, 
or 0.75 for our fiducial values. If we took $z_2\to z_1$ we 
would get $\beta\to1$, while if $z_2\gg z_1=2z$ then 
$\beta\to0.5$.

\begin{figure}[h]
\centering 
\includegraphics[width=\columnwidth]{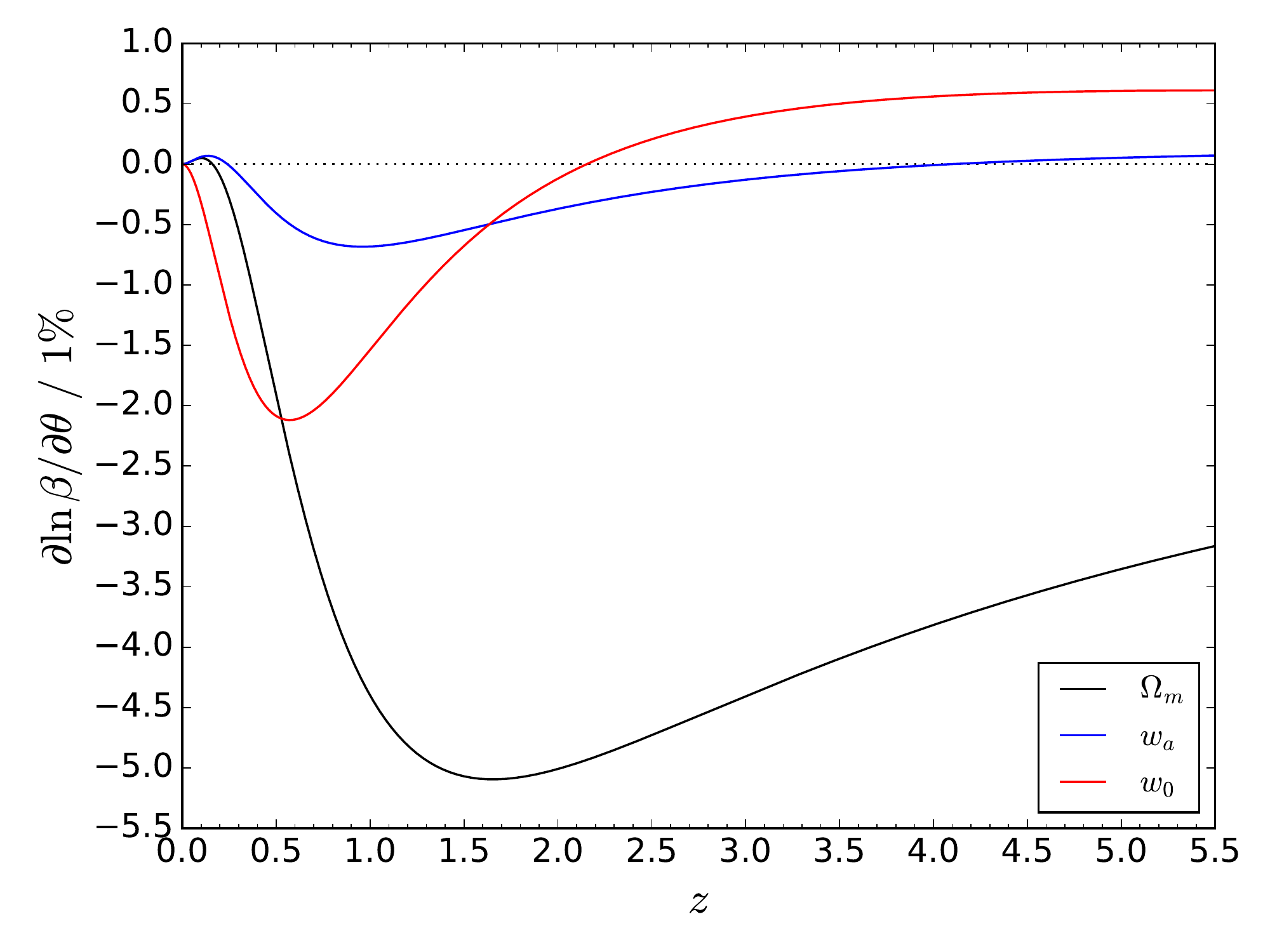}
\caption{The sensitivity of measurements of the double source lensing distance ratio $\beta$ for constraining cosmological parameters $\theta$ 
is plotted as a function of the lens redshift $z$. The magnitude of the sensitivity is here for a 1\% measurement of $\beta$, but the more interesting aspects come from the shape of the curves: the null of the $\Omega_{m}$ curve at $z \approx 0.15$ and the opposite signs for $w_0$ and $w_a$ sensitivities for $z \le 0.23$, as well as $w_a$ becoming 
more sensitive than $w_0$ at $z\approx1.6$, and the null of the $w_0$ 
curve at $z\approx2.1$, indicating distinct behaviors from single 
distance probes.} 
\label{fig:sens} 
\end{figure}

Figure~\ref{fig:sens} presents the cosmological parameter 
sensitivities, following \cite{linder16} but extending the 
results to much higher redshift than considered there. This 
shows several new interesting properties. Between $z\approx1.6$--2.5 
the $\beta$ observable has greater sensitivity to $w_a$ than to 
$w_0$ -- highly unusual among cosmological probes. At $z\approx2.1$, 
there is a null to the influence of $w_0$, which could potentially 
relieve covariance between parameters. As the shapes of the 
sensitivity curves differ between parameters, we expect 
high redshift measurements in general to aid in breaking 
covariances.

\section{Cosmological Leverage of DSPL} \label{sec:constraints} 

The information matrix formalism presents an efficient method 
for combining the sensitivities, taking into account their 
covariances, and the measurement uncertainties, to obtain 
cosmological parameter constraints. We will initially focus on the dark energy 
equation of state space, $w_0$--$w_a$, marginalizing over the 
matter density. To begin with, we consider how observations at 
different redshifts affect the constraints. 

\begin{figure}[h]
\centering 
\includegraphics[width=\columnwidth]{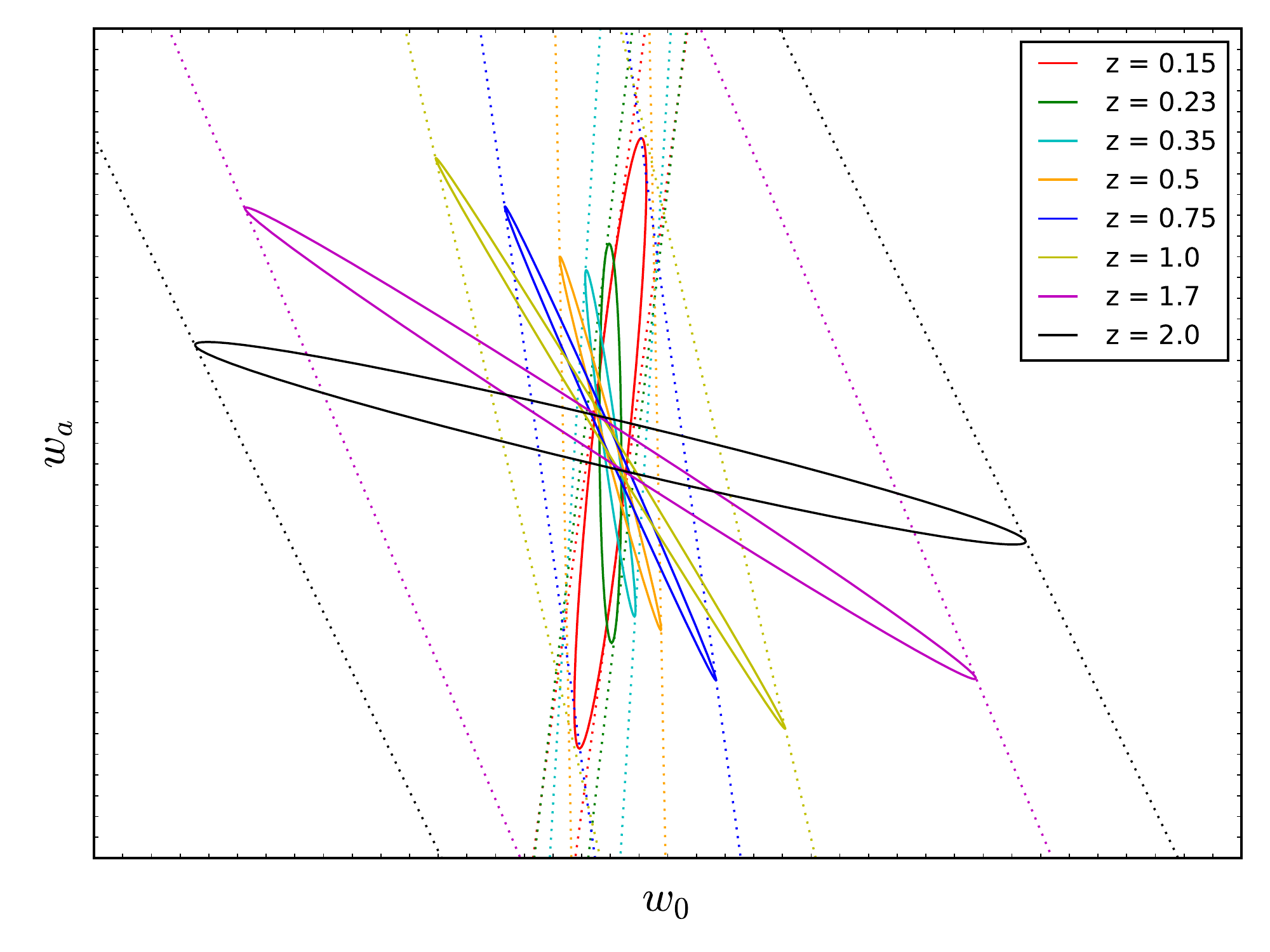}
\caption{The leverage of measurements of the double source lensing distance ratio $\beta$ for constraining the dark energy equation of state value today $w_{0}$ and time variation $w_{a}$ is plotted for 
observations focused at different lens redshifts $z$. Note the 
rotation of the covariance direction, indicating good complementarity 
over a range of redshifts. Solid ellipses fix $\om = 0.3$ for 
clarity while dotted ellipses (extending off the plot) marginalize over $\om$. 
Since the focus is on covariance direction near a single redshift, 
we omit the scale due to idealized precision. 
} 
\label{fig:flower} 
\end{figure}

Figure~\ref{fig:flower} shows that the covariance direction of 
the constraints in the $w_0$--$w_a$ space rotates as the lens redshift 
$z$ increases (keeping the relations 
$z_1/z=2$, $z_2/z_1=1.5$). 
This is clearest when fixing $\om$, as shown by the 
solid contours becoming vertical (strong $w_0$ constraints) near 
the $w_a$ sensitivity null at $z\approx0.23$, and horizontal 
(strong $w_a$ constraints) near the $w_0$ sensitivity null at 
$z\approx2.1$. However the steady rotation (and hence 
complementarity between different redshifts) holds when 
marginalizing over $\om$ (as we do throughout the article), 
as shown by the dotted contours. 
In order to obtain closed contours, we take 
three observations clustered around the labeled redshift, i.e.\ 
at $z$, $z\pm0.05$.

We see that higher redshift measurements are expected to have 
good complementarity with lower redshift ones. Thus the 
upcoming generation of high redshift surveys such as Euclid 
can contribute significantly to dark energy constraints through 
the DSPL probe. For detailed constraints, we study three redshift ranges, roughly 
corresponding to three depths of surveys, for $z=[0.1,0.6]$, 
$[0.6,1.1]$, and $[1.1,1.6]$, each range divided into six bins of width 0.1, e.g.\ with bin centers at $z=0.1$, 0.2, \dots, 0.6. While even higher redshifts could 
be useful, using $\zl>1.6$ would correspond to both $\zs\gtrsim3$, 
making observations more difficult and time consuming. 
In each redshift bin we assume 16 DSPL each with $\beta$ measured 
to 1\% (treated statistically, i.e.\ any systematics common 
across systems are below the 1\% level). This corresponds to 
96 DSPL per set, a reasonable ``gold set''  
for upcoming surveys. 

\begin{figure}[h]
\centering 
\includegraphics[width=\columnwidth]{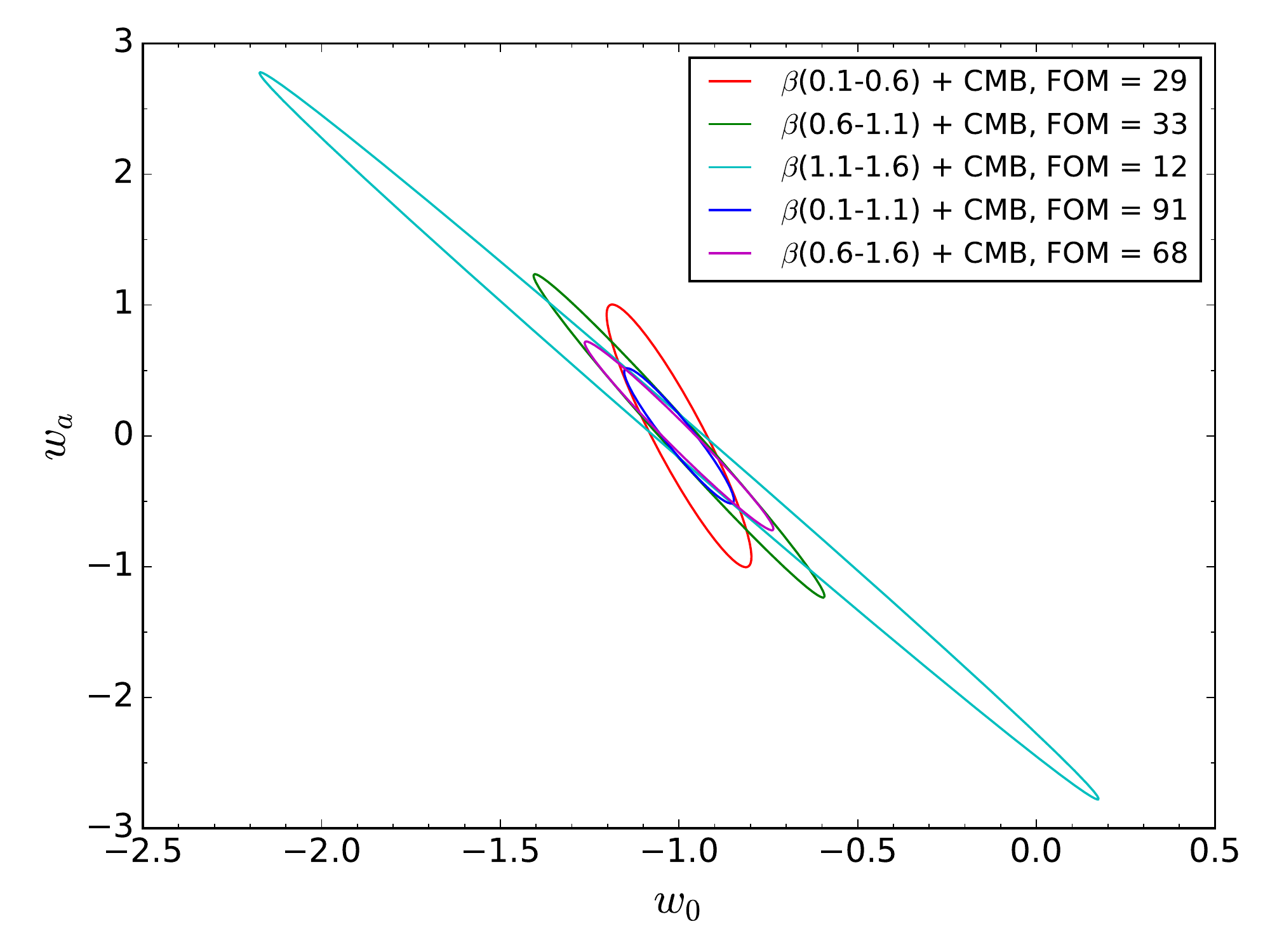}
\caption{1$\sigma$ joint confidence contours on $w_{0}$--$w_{a}$ for DSPL over various redshift ranges, plus CMB. Note the strong 
complementarity of including DSPL all the way out to $z=1.1$. 
} 
\label{fig:dsplcmb} 
\end{figure}

Figure~\ref{fig:dsplcmb} shows the dark energy constraints, 
and figure of merit FOM$=\sqrt{\det F(w_0,w_a)}$, where $F$ is the 
information matrix. We always marginalize over $\om$, and combine  
different redshift ranges of DSPL with external information 
in the form of a Planck prior on the distance to last scattering 
of the cosmic microwave background (CMB). For each individual 
redshift range of DSPL, plus CMB, the dark energy constraints 
are not particularly tight -- this is because the unique virtue 
of DSPL in depending on the higher redshift universe through $D_{ls}$ actually means the constraints 
are weaker in the low redshift range where dark energy dominates. 
However we will shortly see that also including a low redshift 
standard distance probe, such as Type Ia supernova distances 
(SN), will allow the unique leverage of DSPL to work.

The low and middle redshift ranges for DSPL give nearly 
equivalent FOM when combined with CMB. The high redshift range is 
much weaker, since its covariance direction 
(see Fig.~\ref{fig:flower}) is nearly the same as that for CMB. 
Again, the situation will change significantly when we later 
add a standard distance probe as well. Combining complementarity 
redshift ranges for DSPL indeed has a strong effect: for the 
low+mid redshift combination, FOM increases by a factor 3, 
while mid+high redshift gives a factor 2 increase (again not as strong due to 
overlap in covariance direction with CMB). 

Now let us add supernovae (one could equally 
well use distances from baryon acoustic 
oscillations). We use a moderate projected sample (same as 
in \cite{linder16}), with SN concentrated at $z<1$, specifically 
150 local ($z<0.1$), 900 between $z=0.1$--1, and 42 over 
$z=1$--1.7. While Euclid does not include a SN survey 
(but see \cite{desire}), LSST 
will obtain many at $z\lesssim1$, though without spectroscopy; 
the 900 used can be thought of as systematics dominated in 
the SN magnitude measurement, at 
$dm=0.02(1+z)/2.7$ mag; we marginalize over 
the SN effective absolute magnitude $\mathcal{M}$. 
As mentioned above, the inclusion of 
a standard distance probe giving just $D(z)$ enables the 
leverage of DSPL on $D_{ls}(z,z')$ to have great effect. 

\begin{figure}[h]
\centering 
\includegraphics[width=\columnwidth]{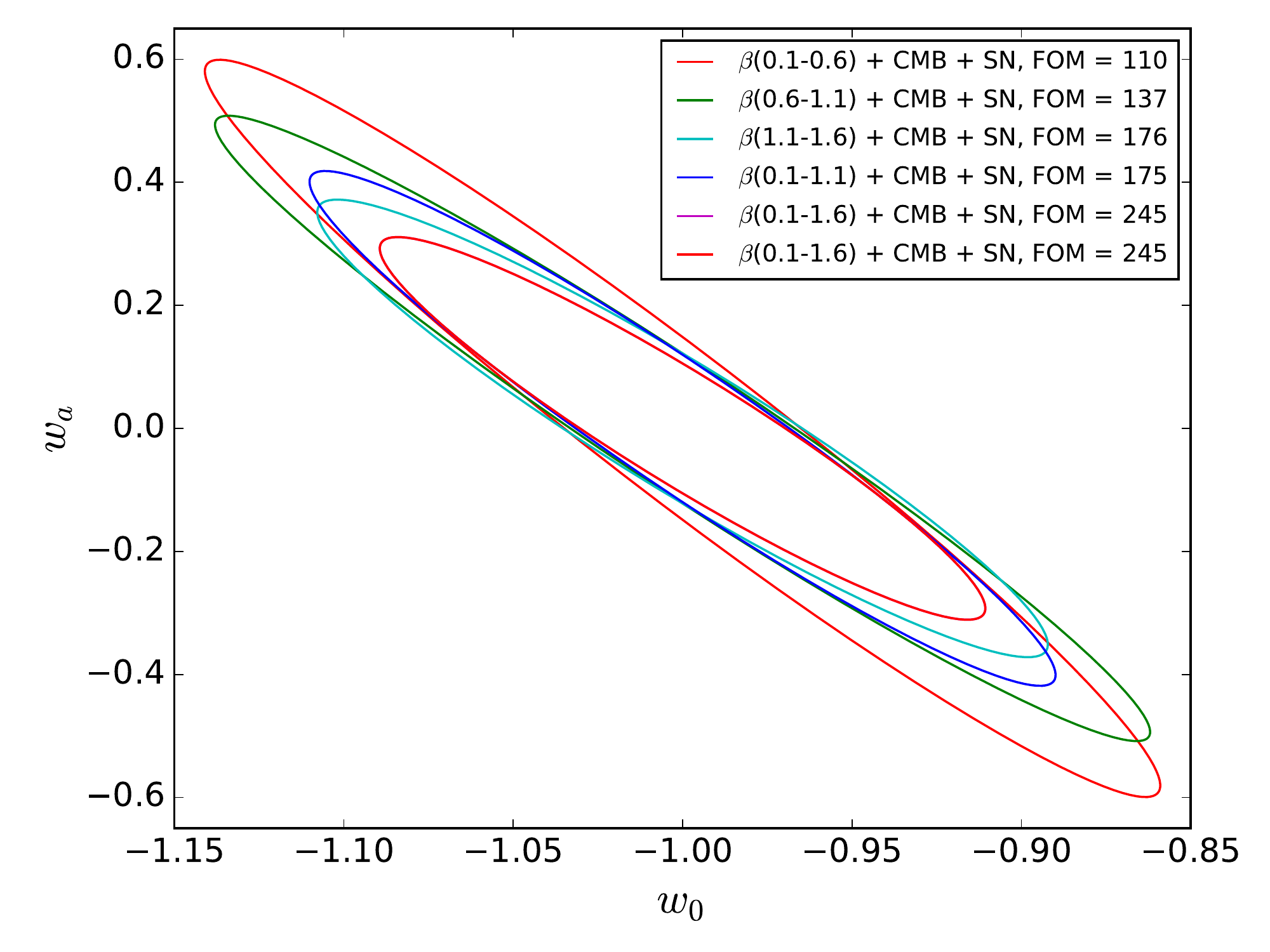}
\caption{1$\sigma$ joint confidence contours on $w_{0}$--$w_{a}$ for DSPL over various redshift ranges, plus CMB and SN. 
Note the complementarity of DSPL with both CMB and SN, leading to increased  
figures of merit.} 
\label{fig:dsplsn} 
\end{figure}

Figure~\ref{fig:dsplsn} displays the cosmological constraints 
from DSPL measurements over various redshift ranges, plus 
combinations of ranges, when including both CMB and SN. Now the 
high redshift set of DSPL gives the best constraints, with FOM=176, 
a factor 15 improvement over without SN. By contrast the 
low and mid redshift DSPL cases improve by a factor $\sim4$. 
When combining low and mid redshift DSPL (and CMB), SN still 
adds an improvement of a factor 1.9 over the case without SN from 
Fig.~\ref{fig:dsplcmb}. All three DSPL redshift ranges 
(so 256 systems total, still a reasonable number) would give FOM=245, compared to 
FOM=72 from CMB+SN without DSPL, i.e.\ 
a factor 3.4 improvement. 
The 1$\sigma$ marginalized uncertainties for 
the case $\beta(0.1-1.6)$+CMB+SN are $\sigma(\Omega_{m})$ = 0.0058, $\sigma(w_0)$ = 0.059, $\sigma(w_a)$ = 0.20.

\section{Source Redshift Distribution} \label{sec:sourcez} 

To check the robustness of the results, we revisit variation of 
the relations $z_1/z=2$ and $z_2/z_1=1.5$. We compute the effects 
on the dark energy FOM as a function of these ratios over all 
lens redshifts, allowing the ranges $z_1/z=[1.1,3]$ and $z_2/z_1=[1.1,3]$. 
The second source redshift $z_2$ however is not allowed to exceed 
$z_2=5$, due to the difficulty in finding such systems owing to 
faintness and reduced galaxy formation rate. 

\begin{figure}[h]
\centering 
\includegraphics[width=\columnwidth]{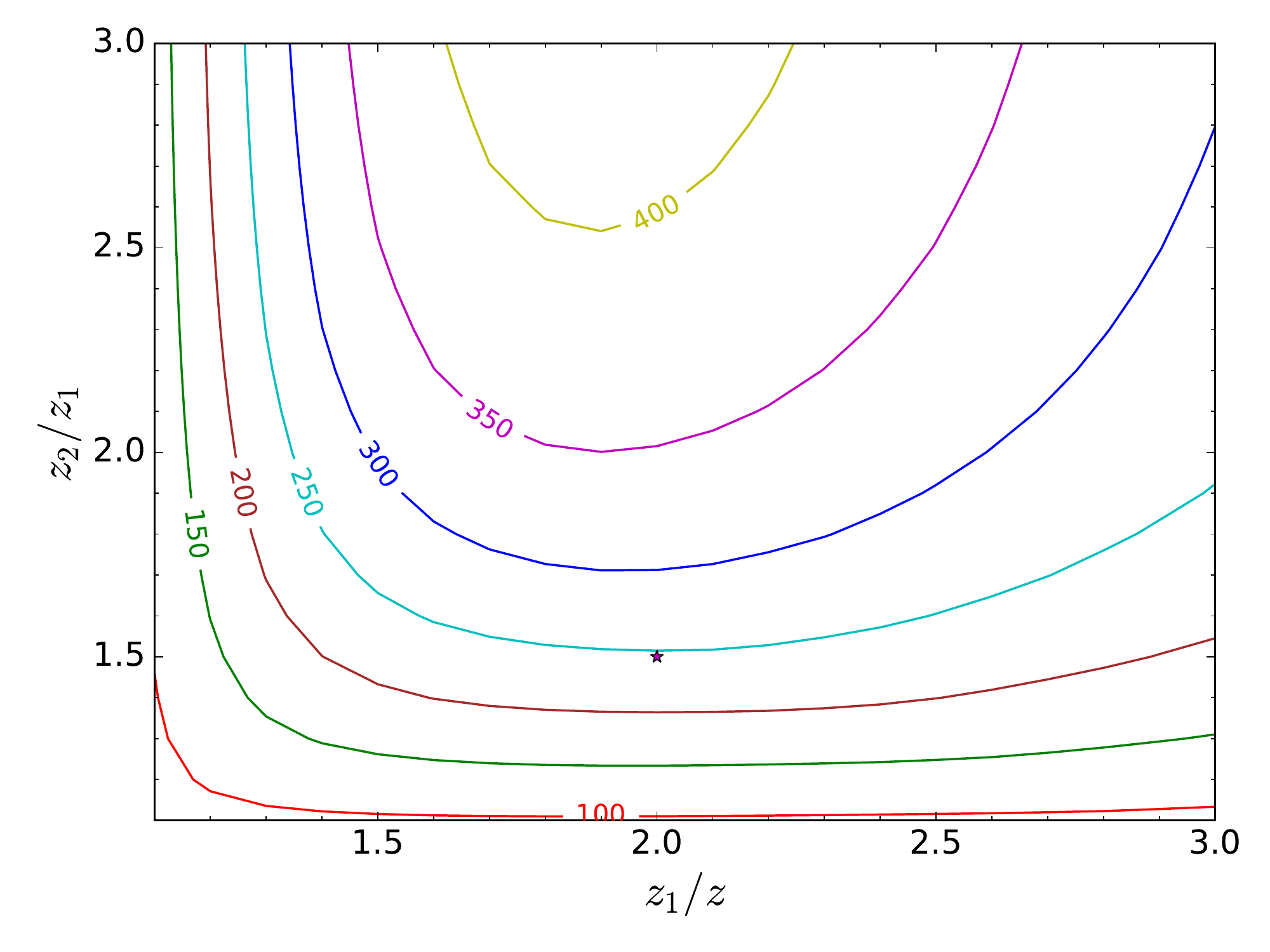}
\caption{FOM isocontours for $\beta(0.1-1.6)$+CMB+SN, with a 
$z_2\le5$ cut imposed. The star in the figure represents the fiducial ratios $z_1/z=2$, $z_2/z_1=1.5$, giving FOM = 245 as in 
Fig.~\ref{fig:dsplsn}. 
} 
\label{fig:isocontour} 
\end{figure}

Figure~\ref{fig:isocontour} shows contours of FOM in the 
$z_1/z$ -- $z_2/z_1$ plane, for the combination of data sets 
that in Fig.~\ref{fig:dsplsn} gave FOM=245: 
$\beta(0.1-1.6)$+CMB+SN. Variation of $z_1/z$ within the 
range 1.5--2.5 has a rather modest effect, changing the 
FOM by less than 10\%, while even $z_1/z=3$ only affects FOM 
at the 20\% level. For $z_2/z_1$, our fiducial value is quite 
a conservative choice, with $z_2/z_1=2$ (2.5) improving FOM by 
40\% (60\%), raising FOM over 340. 
Thus DSPL can be 
a significant contributor to probing the nature of dark energy.


\section{Exploring High Redshift Dark Energy Density} \label{sec:bins} 

Advantageous characteristics of DSPL as a cosmic probe 
include the relatively good sensitivity at high redshift and 
the capability to explore the expansion at redshifts between 
the lens and source redshifts through $D_{ls}$, rather than all 
the way from the observer including the local universe. 
As well, $D_{ls}$ gives the benefit of 
complementarity with standard $D(z)$ probes. Therefore we investigate 
what DSPL can tell us about high redshift dark energy, beyond the 
usual $w_0$--$w_a$ parametrization. 

In this section dark energy density is allowed to float freely 
within high redshift bins, to see how the data can constrain 
dark energy at the epochs when it is predicted to be at the 1--20\% level of 
the critical energy density within the $\Lambda$CDM model. That 
is, we take as parameters $\om$, $\{\ode(z_i)\}$, employing 
five bins with $z_i$ being the centers of $z=[1.1,1.4]$, [1.4,1.7], 
[1.7,2], [2,2.5], [2.5,5]. See also \cite{2106.09581,2106.09713} 
for other probes constraining binned high redshift dark energy 
density. 

\begin{figure}[h]
\centering 
\includegraphics[width=\columnwidth]{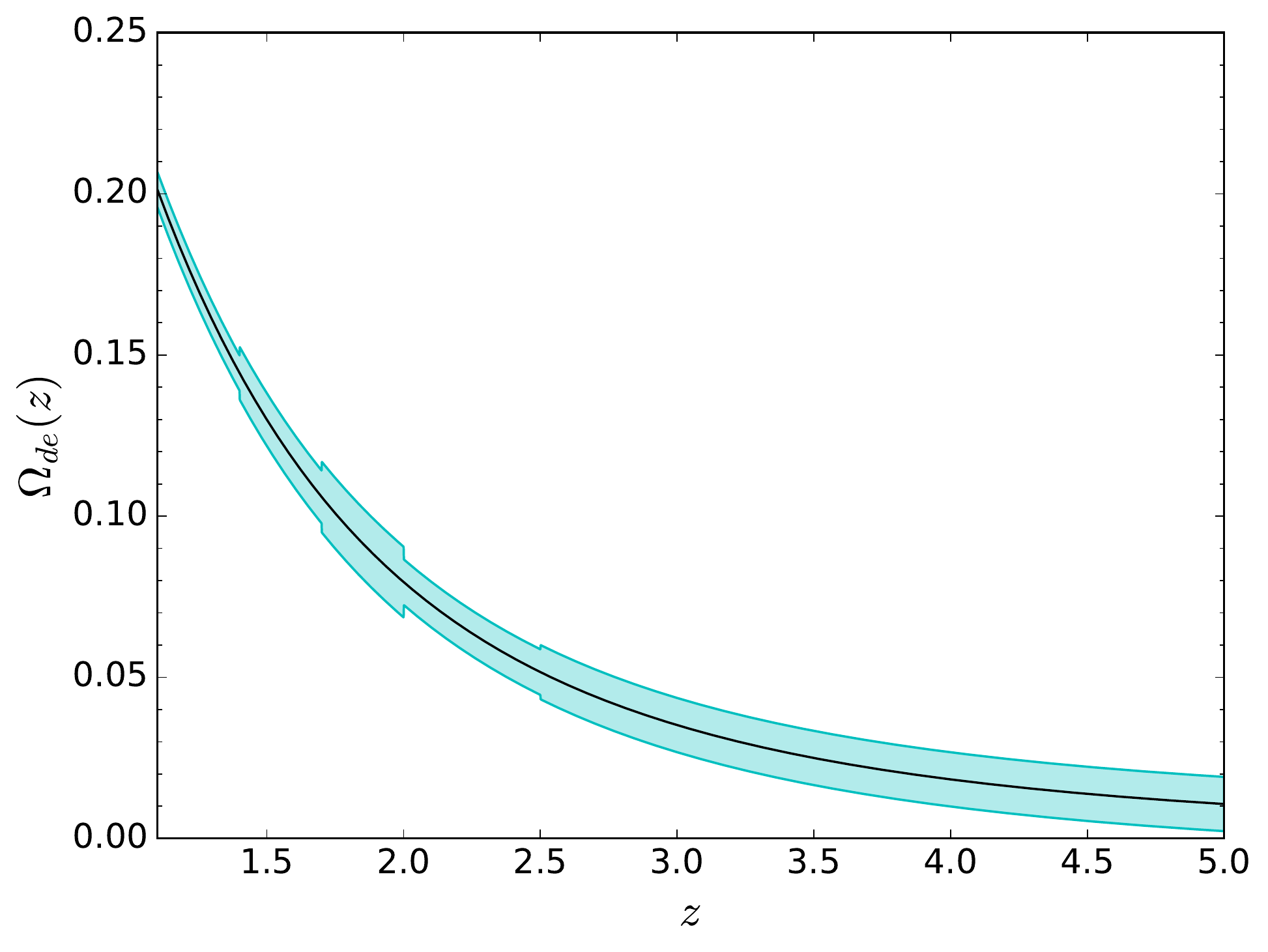}
\caption{Constraints on high redshift dark energy density enabled 
by observations of $\beta(0.1-1.6)$+CMB+SN are shown as a shaded 
band (68\% CL) around the \lcdm\ fiducial behavior (black curve). The presence 
of dark energy could be detected out to $z\approx5$ (at 68\% CL). 
} 
\label{fig:binband} 
\end{figure}

We employ the combined data set as in Fig.~\ref{fig:dsplsn} and 
Fig.~\ref{fig:isocontour}: $\beta(0.1-1.6)$+CMB+SN. 
Figure~\ref{fig:binband} shows the $1\sigma$ marginalized uncertainty 
band on the dark energy density as a function of redshift, across 
the five bins. 
We see that the uncertainty band is distinct from zero dark energy 
density out to $z\approx5$ (at 68\% CL). The magnitudes of the $1\sigma$ marginalized uncertainties are 
$\sigma(\om)=0.0028$, $\sigma(\ode(z_i))=0.0055$, 0.0082, 0.011, 
0.0071, 0.0084 respectively. This would correspond to 31, 
15, 8.4, 9.0, $2.5\sigma$ 
evidence for dark energy at $z=1.25$, 1.55, 1.85, 2.25, 3.75 
respectively. 
(The constraints weaken for bins at higher redshift as dark energy 
is less dynamically important there, then strengthen in the 
last two bins that we chose to be broader.)

\begin{figure*}[h]
\centering 
\includegraphics[width=0.96\textwidth]{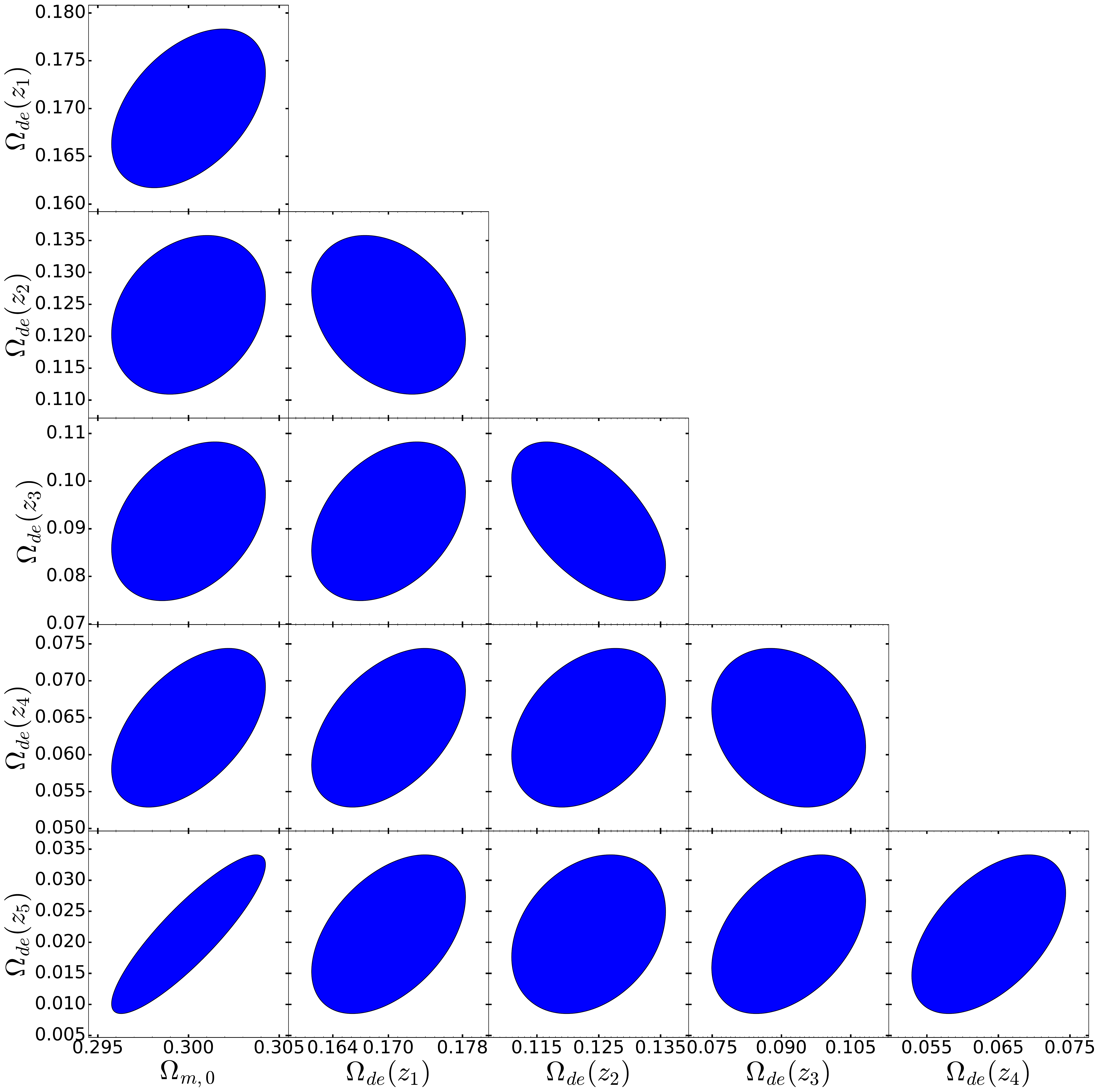}
\caption{2D joint 68\% CL constraints on high redshift dark 
energy density and the present matter density enabled 
by observations of $\beta(0.1-1.6)$+CMB+SN indicate the power 
and complementarity of DSPL for probing dark energy even at 
high redshift.   
} 
\label{fig:bincorner} 
\end{figure*}

Figure~\ref{fig:bincorner} presents a corner plot of the 
2D joint confidence contours for the high redshift binned 
dark energy density parameters, plus the present matter density. 
The combination of data breaks degeneracies significantly, as seen 
by the substantially circular contours, leaving 
the greatest correlation coefficient as $0.88$ between the present 
matter density and the dark energy density in the highest ($z_5$) bin. 
Thus the combination of DSPL, involving $D_{ls}$, and standard 
distance measures $D(z)$ such as from supernovae (or baryon 
acoustic oscillations), plus CMB, is a powerful probe of dark 
energy in the high redshift universe as well.

\section{Conclusions} \label{sec:concl} 

Additional methods for probing cosmology and the nature of 
dark energy to complement and enhance the standard techniques 
would be highly valuable. Double source plane lensing offers 
several promising characteristics, including hundreds of 
expected detections and measurements from the Euclid satellite 
and other surveys, intriguing dependence on the ``remote'' 
distance between lens and source without local universe 
dependence, and strong complementarity between low and high 
redshift observations and with standard distance measures. 

We have quantified the cosmological leverage of DSPL in 
terms of both constraints on dark energy equation of state 
parameters $w_0$, $w_a$, and figure of merit and on freely 
varying binned dark energy density at high redshift. The 
first demonstrates that DSPL, together with moderate level 
CMB and supernovae data, can give FOM $\approx250$, rising 
to $\approx350$ for a less conservative source redshift 
distribution. The second shows that DSPL can be a superb 
probe of the high redshift universe, detecting nonzero dark 
energy density out to $z\approx5$ and giving several 
statistically significant measures of dark energy in independent 
redshift bins between $z\approx1.1$--5. 

Complementarity between cosmic probes -- to break degeneracies, 
crosscheck results, and guard against systematics -- is valuable, 
between $D_{ls}$ and $D(z)$, between low and high 
redshift, and between DSPL and strong gravitational 
lensing time delays. Strong gravitational lensing should 
become a significant, mature technique with the upcoming 
generation of wide surveys, and the extension to the 
$z\gtrsim2$ universe with Euclid 
and future instruments adds a new, further frontier. 

These are exciting prospects, and upcoming surveys should 
keep DSPL as a science case as they develop detection pipelines, 
assess the numbers predicted by \cite{23inLin2016,26inLin2016,27inLin2016,lenspop,1803.03604,oh,2010.15173}, 
and carry out observations. High redshift spectroscopic instruments 
such as MegaMapper will play a critical role in measuring source 
redshifts and for modeling the lens mass profile to see its 
residual impact on the $\beta$ distance ratio. 
Overall, DSPL could provide an important addition to methods 
for understanding the cosmic expansion history.

\acknowledgments  

We thank Joel Brownstein, Leonidas Moustakas, and 
Michael Talbot for assembling the Main Lens Database 
\cite{brownstein} we used to generate Fig.~\ref{fig:knownlenses}. 
We thank Xiaosheng Huang for helpful discussions. 
This work is supported in part by the Energetic Cosmos Laboratory, 
by NASA ROSES grant 12-EUCLID12-0004, and by the 
U.S.\ Department of Energy, Office of Science, Office of High Energy 
Physics, under contract no.\ DE-AC02-05CH11231.

\end{document}